\documentclass[twocolumn,prl,superscriptaddress]{revtex4-1}
\usepackage[margin=1.5cm]{geometry}                
\usepackage{graphicx}
\graphicspath{{./figures/}}
\usepackage{amssymb,amsmath}
\usepackage{epstopdf}
\usepackage{color}
\usepackage{mathtools} 

\newcommand{\bra}[1]{\ensuremath{\left\langle #1\right|}}   
\newcommand{\ket}[1]{\ensuremath{\left|#1\right\rangle}}   

\begin{document}

\title{Observation of quantum state collapse and revival due to the single-photon Kerr effect}

\author{Gerhard  Kirchmair}
\affiliation{Departments of Physics and Applied Physics, Yale University, New Haven, CT 06511, USA}
\author{Brian Vlastakis}
\affiliation{Departments of Physics and Applied Physics, Yale University, New Haven, CT 06511, USA}
\author{Zaki Leghtas}
\affiliation{INRIA Paris-Rocquencourt, Domaine de Voluceau, B.P. 105, 78153 Le Chesnay Cedex, France}
\author{Simon E. Nigg}
\affiliation{Departments of Physics and Applied Physics, Yale University, New Haven, CT 06511, USA}
\author{Hanhee Paik}
\affiliation{Departments of Physics and Applied Physics, Yale University, New Haven, CT 06511, USA}
\author{Eran Ginossar}
\affiliation{Department of Physics and Advanced Technology Institute, University of Surrey, Guildford, Surrey GU2 7XH, United Kingdom}
\author{Mazyar Mirrahimi}
\affiliation{Departments of Physics and Applied Physics, Yale University, New Haven, CT 06511, USA}
\affiliation{INRIA Paris-Rocquencourt, Domaine de Voluceau, B.P. 105, 78153 Le Chesnay Cedex, France}
\author{Luigi Frunzio}
\affiliation{Departments of Physics and Applied Physics, Yale University, New Haven, CT 06511, USA}
\author{S. M. Girvin}
\affiliation{Departments of Physics and Applied Physics, Yale University, New Haven, CT 06511, USA}
\author{R. J.  Schoelkopf}
\affiliation{Departments of Physics and Applied Physics, Yale University, New Haven, CT 06511, USA}

\date{\today}

\begin{abstract}
Photons are ideal carriers for quantum information as they can have a long coherence time and can be transmitted over long distances. These properties are a consequence of their weak interactions within a nearly linear medium. To create and manipulate nonclassical states of light, however, one requires a strong, nonlinear interaction at the single photon level.  One approach to generate suitable interactions is to couple photons to atoms, as in the strong coupling regime of cavity QED systems~\cite{Wallraff:2004vt,Haroche:2007wd}. In these systems, however, one only indirectly controls the quantum state of the light by manipulating the atoms~\cite{Hofheinz:2009tt}. A direct photon-photon interaction occurs in so-called Kerr media, which typically induce only weak nonlinearity at the cost of significant loss. So far, it has not been possible to reach the single-photon Kerr regime, where the interaction strength between individual photons exceeds the loss rate. Here, using a 3D circuit QED architecture~\cite{Paik:2011hd}, we engineer an artificial Kerr medium which enters this regime and allows the observation of new quantum effects. We realize a Gedankenexperiment proposed by Yurke and Stoler~\cite{Yurke:1988wi}, in which the collapse and revival of a coherent state can be observed. This time evolution is a consequence of the quantization of the light field in the cavity and the nonlinear interaction between individual photons. During this evolution non-classical superpositions of coherent states, i.e. multi-component Schr\"odinger cat states, are formed. We visualize this evolution by measuring the Husimi Q-function and confirm the non-classical properties of these transient states by Wigner tomography. The ability to create and manipulate superpositions of coherent states in such a high quality factor photon mode opens perspectives for combining the physics of continuous variables~\cite{Braunstein:2005bp} with superconducting circuits. The single-photon Kerr effect could be employed in QND measurement of photons~\cite{Grangier:1998uk}, single photon generation~\cite{Peyronel:2012je}, autonomous quantum feedback schemes~\cite{Kerckhoff:2010cn} and quantum logic operations~\cite{Milburn:1989kp}.
\end{abstract}

\maketitle

A material whose refractive index depends on the intensity of the light field is called a Kerr medium. A light beam traveling through such a material acquires a phase shift $\phi_{\mathrm{Kerr}}=K \tau I$~\cite{Saleh:ub} where $I$ is the intensity of the beam, $\tau$ is the interaction time of the light field with the material, and $K$ is the Kerr constant. The Kerr effect is a widely used phenomenon in nonlinear quantum optics and has been successfully employed to generate quadrature and amplitude squeezed states~\cite{Slusher:1985iw}, parametrically convert frequencies~\cite{Franken:1961ky}, and create ultra-fast pulses~\cite{Fisher:1969hr}. In the field of quantum optics with microwave circuits, the direct analog of the Kerr effect is naturally created by the nonlinear inductance of a Josephson junction (specifically the $\phi^4 \sim (b+b^\dagger)^4$ term in the Taylor expansion of the $\cos \phi $ of the Josephson energy relation)~\cite{Nigg:2012jj,Bourassa:2012tv}. This effect has been used to create Josephson parametric amplifiers~\cite{Bergeal:2010iu,Mallet:2009fr,CastellanosBeltran:2007ji} and to generate squeezing of microwave fields~\cite{Yurke:1988iha}. However, in both the microwave and optical domains, most experiments utilize the Kerr nonlinearity in a semi-classical regime, where the quantization of the light field does not play a crucial role. The Kerr effect for a quantized mode of light with frequency $\omega_c$, can be described by the normal ordered Hamiltonian
$H_{\mathrm{Kerr}} = \hbar\omega_c a^\dagger a  - \hbar\frac{K}{2} a^\dagger a^\dagger a a$
with $K$ the Kerr shift per photon~\cite{Haroche:2007wd,Bourassa:2012tv}. The average phase shift per photon is again given by $\phi_{\mathrm{Kerr}}=K/\kappa$ with $\kappa$ the decay rate of the photon mode.
Typical Kerr effects are so small that they are not visible on the single photon level as $\kappa > K$. Applications which require $K$ much bigger than $\kappa$ include the realization of quantum logic operations~\cite{Milburn:1989kp}, schemes for continuous variable quantum information protocols~\cite{Braunstein:2005bp} and quantum non-demolition measurements of propagating photons~\cite{Grangier:1998uk}. 

\begin{figure}
\includegraphics[width=89mm]{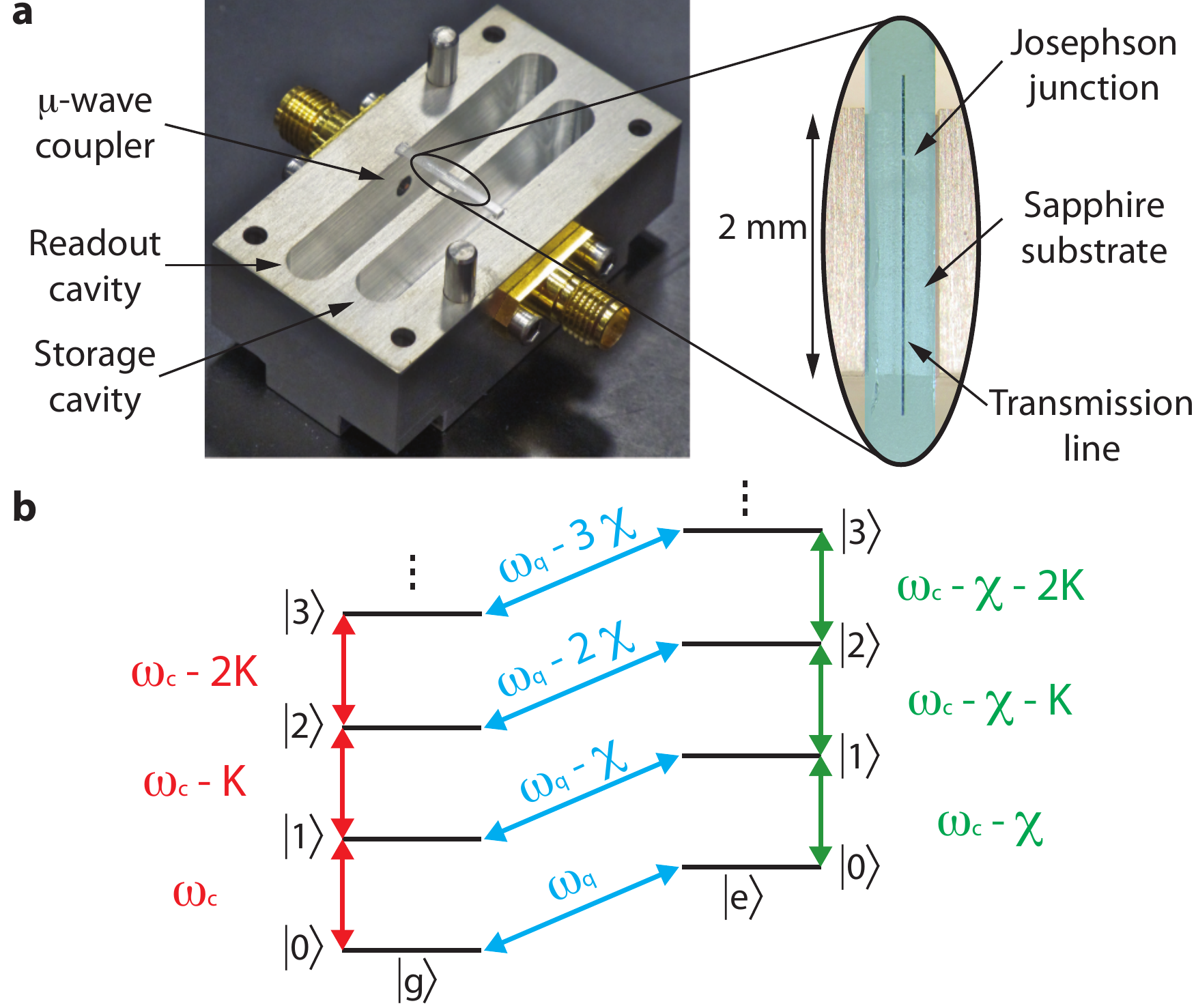} 
\caption{\textbf{Device layout and energy level diagram of the two cavity, one qubit device.} \textbf{a}, Photograph of one half of two aluminum 6061 waveguide cavities coupled to a vertical transmon qubit. The right-hand side of the figure shows a detail of the qubit, fabricated on a c-plane sapphire substrate 1.4 mm wide, 15 mm long, and 430 microns thick. The coupling strength of the qubit is determined by the length of the stripline coupling antenna which extends into each cavity. The upper cavity, with a resonant frequency of $\omega_m/2\pi=$~8.2564~GHz, is used for qubit readout and the lower cavity, with a frequency of $\omega_c/2\pi=$~9.2747~GHz, is used to store and manipulate quantum states. \textbf{b}, Combined energy level diagram of the qubit coupled to the storage cavity. The qubit states  are denoted as \ket{g} and \ket{e} respectively while the cavity states are labeled as \ket{n} with n the number of photons in the cavity. Each photon in the cavity reduces the qubit transition frequency by $\chi$. Equivalently, exciting the qubit reduces the cavity transition frequency by $\chi$. The energy levels of the cavity are not evenly spaced due to the induced Kerr anharmonicity $K/2 \pi=$~325~kHz.}
   \label{fig:device}
\end{figure}

The Kerr nonlinearity of a Josephson junction is also routinely employed to create superconducting qubits. In this case, one engineers a circuit with such a strong anharmonicity that it can be considered as a two level system. By combining a qubit with linear resonators one realizes the analogue of strong coupling cavity QED, known as circuit QED.~\cite{Wallraff:2004vt} This can be used to protect the qubit from spontaneous emission, manipulate and readout its quantum state and couple it to other qubits. One consequence of coupling any resonator to a qubit, is that the resonator always acquires a finite anharmonicity $K$, becoming a Kerr medium itself~\cite{Nigg:2012jj}. In this paper, we have designed $K$ large enough ($K/\kappa > 30$) to be well within the single-photon Kerr regime. At the same time, the nonlinearity is small enough that short  displacement pulses ($t_{\mathrm{pulse}} \approx 10~\mathrm{ns}<< 1 / K$) applied to the resonator create a coherent state, allowing us to conveniently access the large Hilbert space of the oscillator. Only a few experiments have previously come close to the limit where $K/\kappa \approx 1$ while still maintaining the ability to create coherent states~\cite{Yin:2012ib,Hoffman:2011fz,Shalibo:2012wl}.

As a first demonstration within this single-photon Kerr regime, we realize a proposal~\cite{Yurke:1988wi} for creating photonic Schr\"odinger cat states. Specifically, we generate coherent states with a mean photon number of up to four photons and measure the Husimi Q-function of the resonator state using a new experimental measurement protocol. We then show the high quality of the Kerr resonator by measuring the time evolution of the collapse and revival of a coherent state. During the evolution, highly non-classical superpositions of coherent states, i.e. multi-component Schr\"odinger cat states, are formed, which show the coherent nature of the effect. This revival, in contrast to the Jaynes-Cummings revival of Rabi oscillations of a qubit induced by a coherent state~\cite{Haroche:2007wd,Meekhof:1996}, is the revival of a coherent state in a resonator. An analogous effect was indirectly observed in early experiments with a condensate of bosonic atoms in an optical lattice~\cite{Greiner:2002a}.  Additionally, we confirm the nonclassical properties of the transient states by performing Wigner tomography.

We experimentally realize a highly coherent Kerr medium by coupling a superconducting ``vertical'' transmon qubit to two 3D waveguide cavities as shown in Fig.~1a. This design is based on a recently developed 3D circuit QED architecture~\cite{Paik:2011hd}. The two halves of the cavities are machined out of a block of superconducting aluminium (alloy 6061 T6). Both cavities have a total quality factor of about 1 million, corresponding to a single photon decay rate $\kappa/2\pi=$~10~kHz. The vertical transmon consists of a single Josephson junction embedded in a transmission line structure which couples the junction to both cavities. The observed transition frequency of the qubit is $\omega_q/2\pi=$~7.8503~GHz and its anharmonicity is $K_q/2\pi=(\omega_{\mathrm{ge}}-\omega_{\mathrm{ef}})/2 \pi=$~73.4~MHz using the standard convention for labeling from lowest to highest energy level in the qubit as (g,e,f,h,...). The energy relaxation time of the qubit is $T_1= 10~\mu$s with a Ramsey time $T_2^*=8~\mu$s. The qubit is used to interrogate the state of the storage cavity which acts as a Kerr medium. The other cavity is used to read out the state of the qubit after the interrogation, similar to ref.~\cite{Johnson:2010gu}. 

The analysis of the distributed stripline elements and the cavity electrodynamics can be performed using finite-element calculations for the actual geometry. Combined with ``Black-Box'' circuit quantization,~\cite{Nigg:2012jj} one can derive dressed frequencies, couplings, and anharmonicities with good relative accuracy (see supplementary material). For the purposes of the experiments discussed here, the coupling of the qubit to the storage resonator, in the strong dispersive limit of circuit QED, is well described by the Hamiltonian 
\begin{align}
\frac{H}{\hbar} =  \frac{\omega_{q} }{2} \sigma_z - \frac{\chi}{2}a ^\dagger a \sigma_z +  \left(\omega_c -\frac{\chi}{2}\right)a ^\dagger a - \frac{K}{2} a ^\dagger a ^\dagger a a
\label{eqn:JC_approx}
\end{align}
taking into account only the lowest two energy levels of the qubit. The operators $a^\dagger/a$ are the usual raising/lowering operators for the harmonic oscillator and $\sigma_z$ is the Pauli operator. In this description, we completely omit the measurement cavity as it is only used for reading out the state of the qubit and otherwise stays in its ground state. The energy level diagram described by the Hamiltonian given in Eqn.~\ref{eqn:JC_approx} can be seen in Fig.~1b. The second term in Eqn.~\ref{eqn:JC_approx} is the state-dependent shift per photon $\chi/2\pi =$~9.4~MHz  of the qubit transition frequency. The last two terms describe the cavity as an anharmonic oscillator with a dressed resonance frequency $\omega_c$ and a nonlinearity $K/2 \pi=$~325~kHz which is given by $K \approx \chi^2 / 4 K_q$~\cite{Nigg:2012jj}. All interaction strengths in the above Hamiltonian are at least one order of magnitude bigger than any decoherence rate in the system.

To visualize and understand the evolution of the resonator state, we measure the Husimi Q-function $Q_0$ in a space spanned by the expectation value of the dimensionless field quadratures $\mathrm{Re}\{\alpha\}$  and $\mathrm{Im}\{\alpha\}$. $Q_0$ is defined as the modulus squared of the overlap of the resonator state $\ket{\Psi}$ with a coherent state $\ket{\alpha}$ by $Q_0(\alpha)= \frac{1}{\pi}|\left<\alpha|\Psi\right>|^2$. Alternatively, we can write $Q_0$ using the displacement operator $D_\alpha=e^{\alpha a^\dagger-\alpha^* a}$ (note that $D_\alpha^\dagger=D_{-\alpha})$ as $Q_0(\alpha)= \frac{1}{\pi}|\left<0|D_{-\alpha}|\Psi\right>|^2$ which describes the actual measurement procedure employed in the experiment. The sequence to measure $Q_0$ can be seen in Fig.~2a. The initial displacement, $D_\beta$, creates a coherent state $\ket{\Psi}=\ket{\beta}$ in the cavity, whose $Q_0$ is given by a Gaussian, $\frac{1}{\pi} e^{-|\alpha-\beta|^2}$. After a variable waiting time t, we measure $Q_0(\alpha)$ by displacing the cavity state by $-\alpha$ and determine the overlap of the resulting wavefunction with the cavity ground state. The population of the cavity ground state can be measured by applying a photon number state selective $\pi$ pulse, $X^{n=0}_\pi$, on the qubit (see supplementary material). The qubit is excited if and only if the cavity is in the $n=0$ Fock state after the analysis displacement. Applying $\pi$ pulses to the qubit conditioned on other photon numbers, $X^n_\pi$, allows us to measure the overlap of the displaced state with any Fock state n, which we will call the generalized Q-functions $Q_n(\alpha)= \frac{1}{\pi}|\left<n|D_{-\alpha}|\Psi\right>|^2$. In essence we can ask the question: ``Are there n photons in the resonator?'', using photon number state selective pulses~\cite{Johnson:2010gu}. To test the analysis protocol we measured $Q_0$ and $Q_1$ of the cavity in the ground state, Fig.~2b-e, by omitting the first displacement pulse of the sequence given in Fig.~2a. 

\begin{figure}
   \includegraphics[width=89mm]{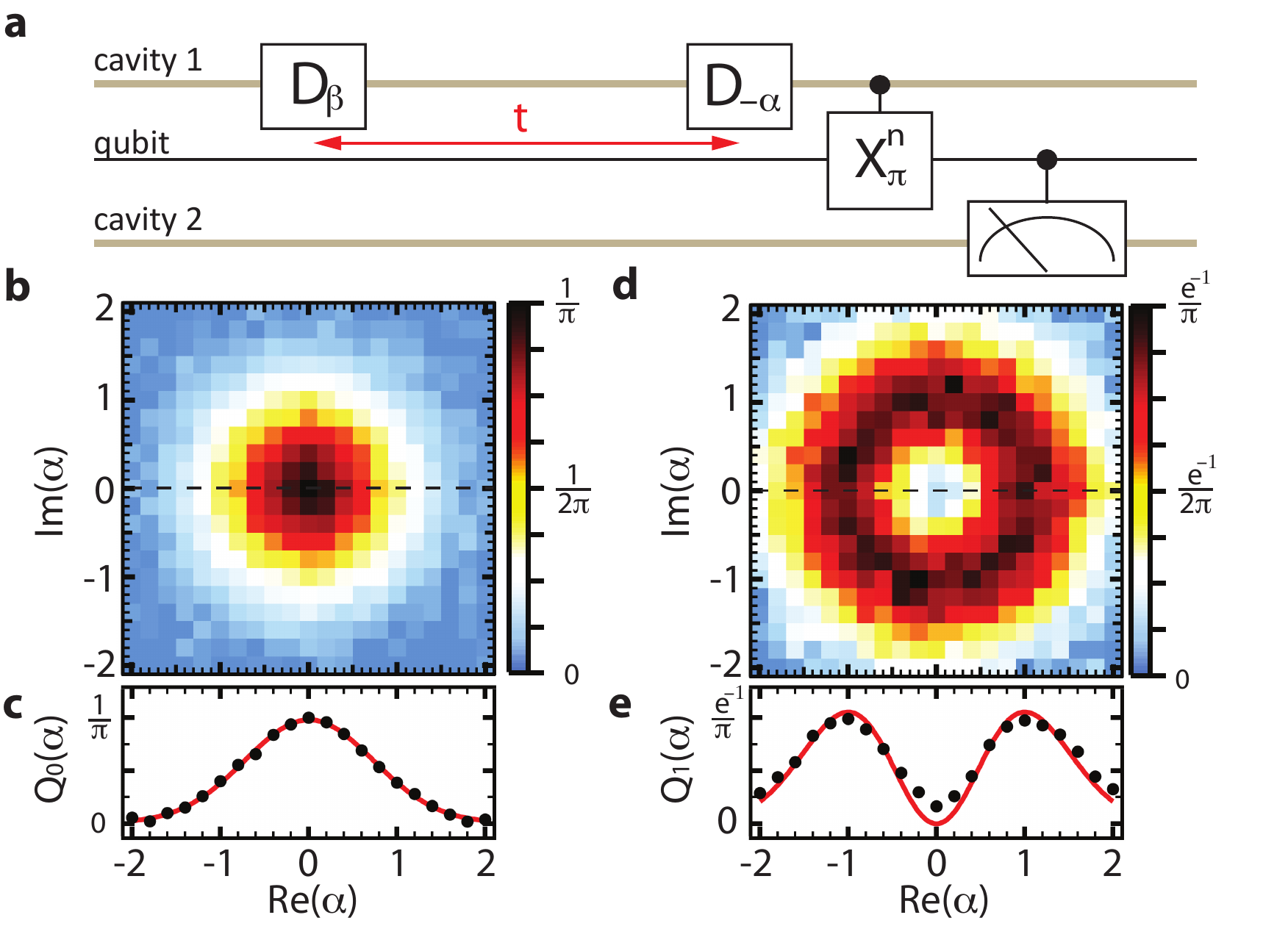} 
   \caption{\textbf{Technique for measuring the generalized Husimi Q-functions.} \textbf{a}, The experimental pulse sequence consists of a 10 ns displacement pulse $D_\beta$ which creates a coherent state in the cavity. After a variable waiting time, t, we analyze the state in the cavity by displacing the cavity by $-\alpha$ followed by a $\pi$ pulse on the qubit conditioned on having n photons in the cavity. In this way we can measure the generalized Q-functions $Q_n(\alpha)= \frac{1}{\pi}|\left<n|D_{-\alpha}|\Psi\right>|^2$ which are projections of the displaced wavefunction onto the Fock states $\ket{n}$ where $Q_0(\alpha)$ is the Husimi Q-function. \textbf{b}, Density plot of $Q_0$ of the ground state. We measure the Q-function at 441 different analysis displacements $\alpha$. The $\pi$ pulse on the qubit $X^{n=0}_\pi$ is conditioned on having no photons in the cavity after the analysis displacement. \textbf{c}, Linecut of $Q_0$ of the ground state along $\mathrm{Im}\{\alpha\}=0$. The red line is a plot of the theory, a Gaussian given by $\frac{1}{\pi} e^{-|\alpha|^2}$. A fit to the data with a Gauss function $\frac{1}{\pi}e^{-I^2/2 \sigma^2}$ results in $2 \sigma^2 = 1.03 \pm 0.02$ which is consistent with the expected width. \textbf{d}, Density plot of $Q_1$ of the ground state. In this case the $\pi$ pulse on the qubit $X^{n=1}_\pi$ is conditioned on having one photon in the cavity after the analysis displacement. \textbf{e}, Linecut of $Q_1$ of the ground state along $\mathrm{Im}\{\alpha\}=0$. The red line is a plot of the theory given by a Poisson distribution $\frac{1}{\pi}|\alpha|^2 e^{-|\alpha|^2}$.}
   \label{fig:sequence}
\end{figure}

\begin{figure*}
   \centering
   \includegraphics[width=183mm]{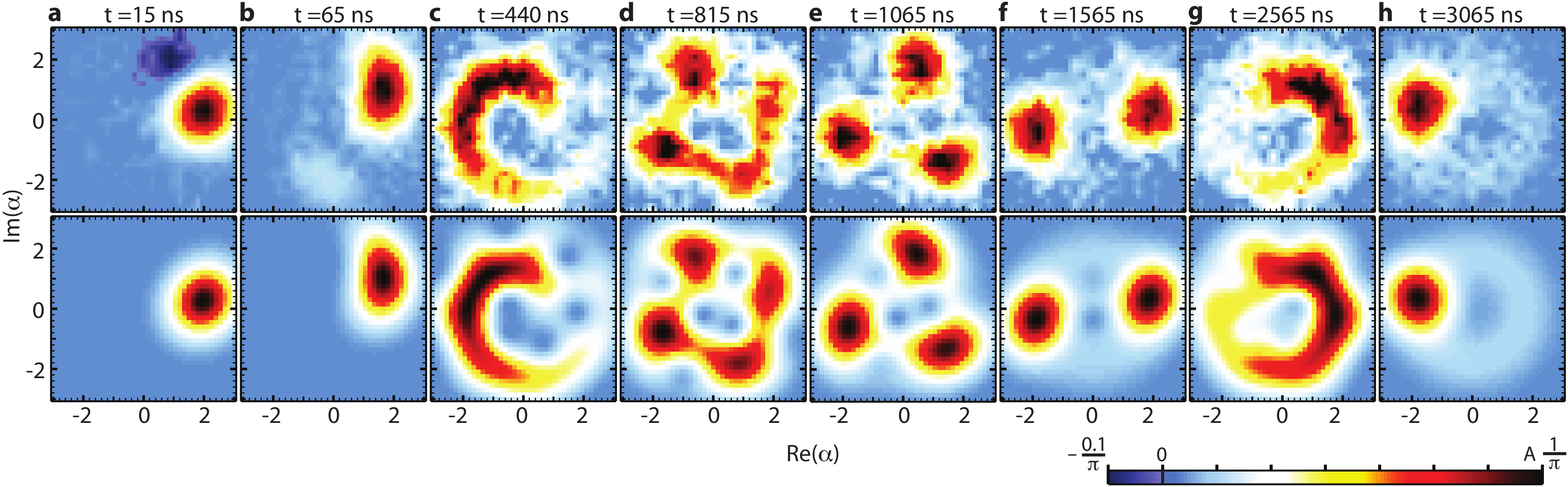} 
   \caption{\textbf{Time evolution of \boldmath{$Q_0$} for a coherent state in the nonlinear cavity.} Experiment (upper row) and theory (lower row) for a coherent state $\beta=$ 2. The time t for the frames \textbf{a}-\textbf{h} is given above each panel. We measure $Q_0$ at 441 different analysis displacements  $\alpha$. The resolution of the pictures was doubled by interpolation. Initially the phase of the state spreads rapidly, \textbf{a}-\textbf{c}, leading to a complete phase collapse after a characteristic time $T_{\mathrm{col}}=\frac{\pi}{2 \sqrt{\bar{n}}K}=$ 385ns. For short times, the Kerr interaction leads to a quadrature squeezed state along the $\mathrm{Re}\{\alpha\}$ axis which can be seen in \textbf{b}. After the complete phase collapse structure emerges again,  \textbf{d}-\textbf{h}, at times which are integer submultiples of the complete revival time $T_{\mathrm{rev}}=\frac{2 \pi}{K}$. At these times, one obtains coherent state superpositions which are multi-component cat states, up to a maximum number of resolvable components set by the average photon number in the initial coherent state displacement. The color scale of each pair of plots is individually rescaled with the scaling parameter given for \textbf{a}-\textbf{h} by A = 1, 0.93, 0.32, 0.24, 0.3, 0.46, 0.24, 0.89. The negative amplitudes in plot \textbf{a} are due to a 10\% excited state population of the qubit which evolves in a different rotating frame (see supplementary material). This excited state population is not visible in the other frames as it disperses quickly and vanishes in the large positive amplitudes.}
\label{fig_Kerr}
\end{figure*}

Using this method, we can follow the time evolution of a coherent state in the presence of the Kerr effect. In the experiment, we prepare a coherent state with an average photon number $|\beta|^2=\bar{n}=4$ using a microwave pulse~\cite{Shalibo:2012wl} to displace the cavity. We then measure $Q_0$ for different delays between the preparation and analysis pulses. A comparison of the theoretical evolution of the coherent state and the measured evolution can be seen in Fig.~3. The time evolution of the state is described by considering the action of the Kerr Hamiltonian $H_{\mathrm{Kerr}}$ on a coherent state $\ket\beta$ in the cavity~\cite{Yurke:1988wi,MILBURN:1986vh}. In the rotating frame of the harmonic oscillator, with the qubit in the ground state, we can write
\begin{equation}\label{eqn:unitary_kerr}
\ket{\Psi(t)}= e^{i \frac{K}{2} (a^\dagger a)^2  t} \ket{\beta}  = e^{-|\beta|^2/2} \sum_n \frac{\beta^n}{\sqrt{n!}} e^{i \frac{K}{2} n^2 t} \ket{n}.
\end{equation}
For short times, the nonlinear phase evolution of the Fock states $\ket{n}$ is closely approximated by a rotation of the state with an angle $\phi_{\mathrm{Kerr}}=K t(|\beta|^2+1/2)$ with respect to the frame rotating at $\omega_c$. The onset of this rotation can be seen in Fig.~3a which is taken at the minimal waiting time of 15 ns between the two displacement pulses. Due to this waiting time, the state rotates under the influence of the Kerr effect from $\beta = 2$ to $\beta e^{i \phi_{\mathrm{Kerr}}} = 2 e^{i 0.13} $. For longer times we can see how the state rotates further and spreads out on a circle, Fig.~3b,c. This spreading can be simply understood in a semi-classical picture where the amplitude components in the coherent state further away from the origin evolve with a higher angular velocity given by the $n^2$ dependence of the Kerr effect. Complete phase collapse is reached at a time when the phase dispersion across the width of the photon number distribution corresponds to $\approx \pi$, which can be estimated as $T_{\mathrm{col}}=\frac{\pi}{2 \sqrt{\bar{n}}K}$~\cite{Haroche:2007wd}. For our system the complete phase collapse happens at 385~ns. Fig.~3b, shows that a Kerr medium can be used as a resource to generate squeezed states~\cite{MILBURN:1986vh}. The state is squeezed along the $\mathrm{Re}\{\alpha\}$ quadrature with a width of 0.88(1) as predicted from theory. The maximum squeezing occurs at $t= 58$~ns with a width of $Q_0$ of 0.87.

After the complete phase collapse, structure re-emerges in the form of multi-component superpositions of coherent states, Fig.~3d-f, at times which are integer submultiples of the complete revival $T_{\mathrm{rev}}=\frac{2 \pi}{K}$, Fig.~3h. The revivals, periodically appearing every $T_{\mathrm{rev}}$, can be understood by noting that $e^{i \frac{K}{2} n^2 t}=(-1)^{n^2} $ for $t=T_{\mathrm{rev}}$. The cavity state is then given by 
$
\ket{\Psi(T_{rev})}= \ket{-\beta}.
$
At this time we get a complete state revival to a coherent state with opposite phase.
For $t=T_{\mathrm{rev}}/q$, with $q$ an integer larger than 1, we can write the state of the oscillator as a superposition of q coherent states~\cite{Haroche:2007wd}
\begin{equation}
\ket{\Psi\left(\frac{T_{\mathrm{rev}}}{q}\right)}=\frac{1}{2 q} \sum^{2q-1}_{p=0}  \sum^{2q-1}_{k=0} e^{i k (k-p)\frac{\pi}{q}}\ket{\beta \; e^{i p \frac{\pi}{q}}}.
\end{equation}
For q=2 we get the two-component Schr\"odinger cat state similar to the cat states created in ref.~\cite{Deleglise:2008wo,Meekhof:1996}.
To distinguish the q components of a cat state, the coherent states have to be separated by more than twice their width on a circle with a radius given by the initial displacement. In other words, the coherent states have to be quasi-orthogonal. This means for a displacement of $|\beta| =2$ the maximum number of coherent states that can be distinguished is 4.

In frame Fig.~3g we can see how the state again completely dephases shortly before the coherent revival in Fig.~3h after $t=3065$ ns. After this time we get a state with amplitude $|\beta|=1.78(2)$ which fits to the expected decay of the resonator state. The theoretical plots for Fig.~3 were simulated by solving a master equation using the decay rate $\kappa/2\pi=10$ kHz of the resonator and introducing a small detuning of 5 kHz of our drive from the resonator frequency $\omega_c$. The time evolution of the state in the experiment agrees well with the theory. The hazy ring that can be seen in theory and experiment is produced by cavity decay. The evolution of the state from 0 - 6.05 $\mu$s over 50 frames, including two revivals,  can be seen in a movie provided in the supplementary material.

To get a more quantitative comparison of experiment and theory we want to determine the quantum state of the resonator by measuring its Wigner function. Although in principle one can reconstruct the cavity wavefunction from the measured Husimi Q-function, in practice there is important information, such as the  interference fringes between coherent state superpositions, which is exponentially suppressed as their separation increases~\cite{Haroche:2007wd}. This makes it hard to distinguish a mixture of coherent states from a coherent superposition. An experimentally more practical way to determine the quantum state of a resonator is to measure its Wigner function as it emphasizes the interference fringes. Previously, the Wigner function of a cavity~\cite{Haroche:2007wd} has been determined by measuring the parity of the resonator state. Here, we use a modified technique based on earlier work with ion traps and microwave circuits~\cite{Hofheinz:2009tt,Leibfried:1996}. The main difference is our ability to directly measure $Q_n(\alpha)$ which allows for a simple and efficient measurement and reconstruction of the density matrix of the resonator by using a least square fit to each $Q_n$ (see supplementary material). Using this density matrix we then calculate and plot the Wigner function.

In Fig.~4 we show a comparison of the experimentally obtained Wigner functions to a simulation at three different times during the state evolution. The times ($t= 2\pi/2K, 2\pi/3K, 2\pi/4K$) were selected such that the Wigner functions correspond to two, three and four-component cat states. The simulation was again done by solving a master equation which includes the decay of the cavity. The fidelity $F=\bra{\Psi_{\mathrm{id}}} \rho_{\mathrm{m}} \ket{\Psi_{\mathrm{id}}}$ of the measured state $\rho_{\mathrm{m}}$, compared to an ideal n-component cat state $\ket{\Psi_{\mathrm{id}}}$, consisting of coherent states with amplitude $|\beta|=2 e^{-\kappa t/2}$, is $F_2=0.71, F_3=0.70, F_4=0.71$ for the two, three, four-component cats respectively.  The Wigner functions show clear interference fringes which demonstrates that the evolution is indeed coherent and well described by the wavefunction given in Eqn.~\ref{eqn:unitary_kerr}, up to the decay of the cavity. The main reduction in the fidelity is due to the spurious excited state population of the qubit (see supplementary material) and the decay of the resonator. The decay of the resonator state is also responsible for the asymmetry in the interference fringes of the Wigner function, e.g. the maximum of the interference fringes for the two-component cat states is shifted to the left in both theory and experiment.

\begin{figure}   
\label{fig:cats}
\includegraphics[width=89mm]{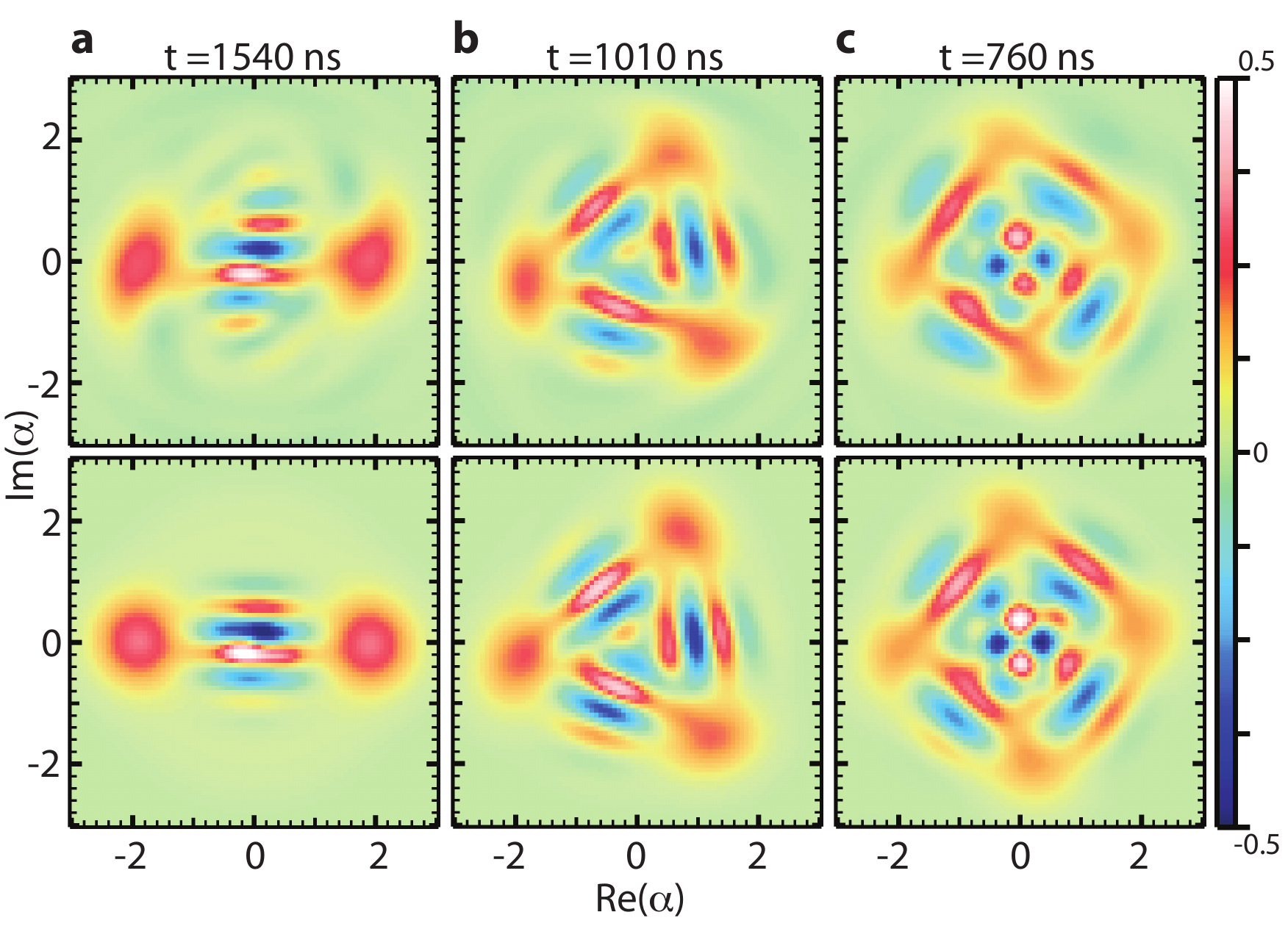} 
\caption{\textbf{Wigner function of the multi-component cat states emerging during the Kerr interaction}. The top row shows the measured Wigner functions of a coherent state subject to a Kerr interaction for a time t. The lower row shows the theoretically expected Wigner functions for the same interaction time obtained by a simulation including the decay of the cavity. The Wigner functions are reconstructed from measurements of the quasi probability distributions $Q_n(\alpha)$ for $n=0-7$. Each $Q_n(\alpha)$ was measured at the same displacements as the corresponding data in Fig.~3. Comparing the experimentally obtained state to an ideal cat state we find a fidelity $F_2=0.71, F_3=0.70, F_4=0.71$  for the two, three, four-component cats respectively. The experimental data show excellent correspondence with theoretical predictions, including the interference fringes and regions of negative quasi-probability distribution, confirming that highly nonclassical states are produced by the Kerr evolution.}
\end{figure}


In conclusion, we have shown that we can engineer strong photon-photon interactions in a cavity, entering the single-photon Kerr regime where $K>> \kappa$. We are able to observe the collapse and revival of a coherent state due to the intensity-dependent dispersion between Fock states in the cavity. This opens the possibility to use such a Kerr medium for error correction schemes where a nonlinear cavity is used to realize the necessary components~\cite{Kerckhoff:2010cn}. The good agreement between the theory and the experiment demonstrates the accurate understanding of this system. It also confirms our ability to predict higher-order couplings which is a necessary ingredient for understanding the behavior of large circuit QED systems. Furthermore, we have measured the evolution of a coherent state in a Kerr medium at the single photon level and shown a new experimental way for creating and measuring multi-component Schr\"odinger cat states. This demonstrates the ability to create, manipulate and visualize coherent states in a larger Hilbert space and opens up new directions for continuous variable quantum computation~\cite{Leghtas:2012vv}.

\begin{acknowledgments}
We thank M.H.Devoret, M.D. Reed, M. Hatridge and A. Sears for valuable discussions. This research was supported by the National Science Foundation(NSF) (PHY-0969725), the Office of the Director of National Intelligence (ODNI), Intelligence Advanced Research Projects Activity (IARPA) through the Army Research Office (W911NF-09-1-0369), and the U.S. Army Research Office (W911NF-09-1-0514). Facilities use was supported by the Yale Institute for Nanoscience and Quantum Engineering (YINQE) and the NSF (MRSECDMR 1119826). SMG acknowledges support from the NSF (DMR-1004406). MM and ZL acknowledge support from French ÒAgence Nationale de la RechercheÓ under the project EPOQ2 (ANR-09-JCJC-0070). SEN acknowledges support from the Swiss NSF. EG acknowledges support from EPSRC (EP/I026231/1).

\end{acknowledgments}

\bibliographystyle{naturemag}
\bibliography{biblio}

\end{document}


\title{Observation of quantum state collapse and revival due to the single-photon Kerr effect}
\author[1]{Gerhard \ Kirchmair}
\author[1]{Brian\ Vlastakis}
\author[2]{Zaki\ Leghtas}
\author[1]{Simon\ E.\ Nigg}
\author[1]{Hanhee\ Paik}
\author[3]{Eran\ Ginossar}
\author[1,2]{Mazyar\ Mirrahimi}
\author[1]{Luigi\ Frunzio}
\author[1]{S.\ M. \ Girvin}
\author[1]{R.\ J. \ Schoelkopf}
\affil[1]{Departments of Physics and Applied Physics, Yale University, New Haven, CT 06511, USA}
\affil[2]{INRIA Paris-Rocquencourt, Domaine de Voluceau, B.P. 105, 78153 Le Chesnay Cedex, France}
\affil[3]{Department of Physics and Advanced Technology Institute, University of Surrey, Guildford, Surrey GU2 7XH, United Kingdom}

\maketitle

\section{Materials and Methods}
\subsection{Qubit fabrication}

Josephson junctions and dipole antennas are fabricated in one single fabrication step by electron-beam lithography followed by aluminum double-angle electron-beam evaporation. The two evaporations deposit thin aluminium films with a thickness of 20 nm and 60 nm respectively. These layers are separated by an AlOx barrier grown by thermal oxidation for 720 seconds in 2000 Pa static pressure of a gaseous mixture of 85\% argon and 15\% oxygen.

\subsection{Measurement setup}
\begin{figure*}[h] 
   \includegraphics[width=183mm]{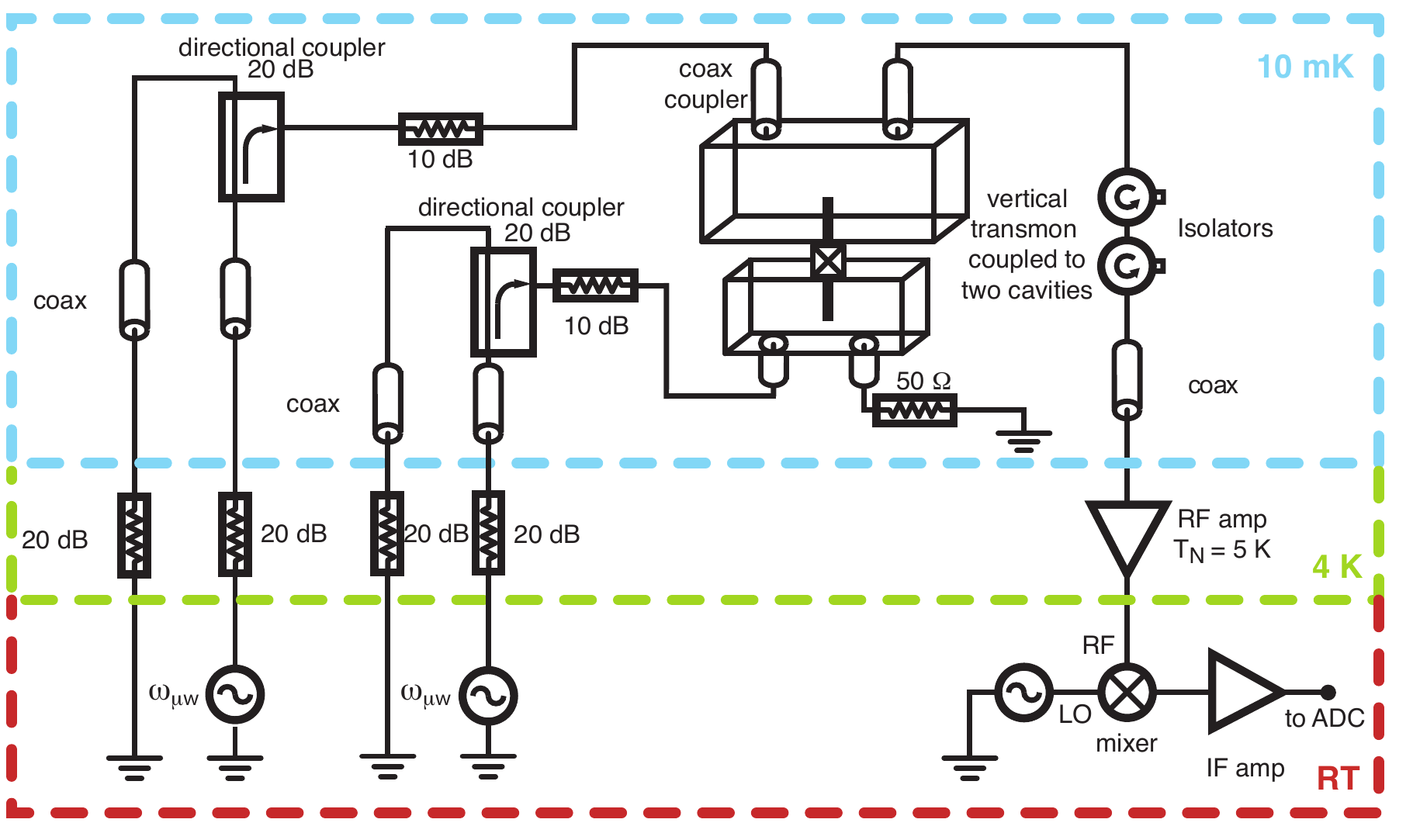} 
   \caption{\textbf{Block diagram of the measurement setup.}}
         \label{fig:setup}
\end{figure*}

The state of the qubit is probed by measuring the qubit-state dependent cavity transmission
using heterodyne detection. Fig.~\ref{fig:setup} is a block diagram of the measurement setup. Microwave
signals for control and measurement transmit through a coax cable with a 20 dB attenuator at 4 K and
a 20 dB directional coupler followed by a 10 dB attenuator at 25 mK. The cavity transmission at
the output goes through two cryogenic isolators at 25 mK followed by a HEMT amplifier
with a noise temperature of 5~K at the 4~K stage. At room temperature, the transmission signal from the cavity is further amplified by two low-noise room temperature amplifiers and is mixed down to a 20~MHz signal which is digitized using a 1~GS analog to digital converter.

\section{Full system Hamiltonian}
The two cavities, one qubit system, can be described by the Hamiltonian
\begin{align}\label{full_eqn}
\frac{H}{\hbar} &=\;  \omega_c a^\dagger a  +\omega_m a_m^\dagger a_m +\omega_q b^\dagger b 
\\&- \frac{K}{2} a^\dagger a^\dagger a a - \frac{K_m}{2} a_m^\dagger a_m^\dagger a_m a_m  - \frac{K_q}{2} b^\dagger b^\dagger b b \nonumber
\\&- \chi a^\dagger a \; b^\dagger b - \chi_{qm} a_m^\dagger a_m \; b^\dagger b - \chi_{cm} a^\dagger a \; a_m^\dagger a_m\nonumber
\end{align}
in the strong dispersive limit of circuit QED. Where $a/ a^\dagger,\; a_m/ a_m^\dagger,\; b/b^\dagger$  the raising/lowering operators for the storage cavity, the measurement cavity and the qubit. The Hamiltonian is calculated by treating the system as three coupled harmonic oscillators and introducing the cosine term in the Josephson relation, using the $\phi^4 \sim (b+b^\dagger)^4$ term in the Taylor expansion, as a perturbation~\cite{Nigg:2012jj}. Higher order terms ($\sim \phi^6$) are more than a factor of 1000 smaller for the parameters of this experiment and can be safely neglected. The Hamiltonian in Eqn.1 of the main text can be recovered from the above Hamiltonian by neglecting all but the lowest two levels of the qubit ($b^\dagger b^\dagger b b = 0$) and replacing $b^\dagger b$ with $\frac{1}{2} (1+\sigma_z)$. Furthermore all terms $\sim a_m^\dagger a_m$ are zero as the readout resonator always stays in the ground state during the Kerr evolution.

The first three terms in the above Hamiltonian describe each mode as a harmonic oscillator with $\omega_c$ the resonance frequency for the storage cavity, $\omega_m$ the resonance frequency for the measurement cavity and $\omega_q$ the ground to first excited state transition frequency for the qubit. The next three terms describe the anharmonicity of each mode, the anharmoncity of the storage resonator $K$, the anharmonicity of the measurement cavity $K_m$ and the anharmonicity of the qubit $K_q$. The last three terms describe the state dependent shifts of each mode due to the state of the other two. This means that the transition frequency of a resonator not only depends on the state of the qubit but also on the state of the other cavity. In our system all of these state dependent shifts are much bigger then any decay rate. The state dependent shift of the storage cavity to the qubit is $\chi$, of the measurement cavity to the qubit is $\chi_{qm}$ and of the storage cavity to the readout cavity $\chi_{cm}$. The values of all parameters determined by spectroscopy and a comparison to the theory values obtained using finite-element calculations for the actual geometry, combined with ``Black-Box'' circuit quantization~\cite{Nigg:2012jj} are given in Table \ref{parameters}.

\begin{table}[h]
\caption{Comparison of the experimentally obtained and predicted values for the frequency and anharmonicity of each mode as well as the state dependent shifts between the modes.}
\begin{center}
\begin{tabular}{c||c|c|c}
& Exp. (MHz)& Th. (MHz)& deviation (\%)\\
\hline
$\omega_q/2\pi$ &7850.3&7890& $<$ 1\\
$\omega_c/2\pi$ &9274.7&9372&1\\
$\omega_m/2\pi$ &8256.4&8336&1\\
$K_q/2\pi$ & 73.4& 72& 2\\
$K/2\pi$ & 0.325& 0.25& 30\\
$K_m/2\pi$ & 3.8& 3.7&3\\
$\chi/2\pi$ &9.4 & 8.2 &15 \\
$\chi_{qm}/2\pi$ & 29.5 & 29.5& $<$ 1\\
$\chi_{cm}/2\pi$ &2.45 & 2.1 &16\\
\end{tabular}
\end{center}
\label{parameters}
\end{table}

\section{Qubit state readout and Cross Kerr dependence}

The state dependent shift between the two cavities, or Cross-Kerr, enables us to readout whether one cavity is in the ground state or not by probing the other cavity exactly like reading out the qubit state. This also means that the qubit cannot be readout independently of the state of the storage cavity. Both qubit and storage cavity in an excited state will contribute to the readout voltage. To remove this effect in our measurements, e.g. determining whether the qubit was excited by a photon number state selective $\pi$ pulse, we always perform a control experiment without a $\pi$ pulse applied to the qubit. This measurement gives us the readout voltage corresponding to the state of the storage cavity only. As our readout is linear, we can subtract this voltage from the combined measurement and thus infer the qubit state. This subtraction was done to measure all $Q_n(\alpha)$ in the main text.

\section{Photon number state selective pulses}

In order to measure the state of the storage cavity, we must project the cavity onto a photon number (Fock) state.  From Eqn.~\ref{full_eqn}, we can see that the qubit transition frequency is dependent on the photon number state of the resonator. As long as the frequency shift, $\chi$, is much larger than spectral width $\sigma$ of the interrogation pulse, a photon number state selective pulse $X_\pi^m$ can be performed. In this experiment, $\chi/2\pi = 9.4$~MHz and $\sigma/2\pi = 2.6$~MHz, making a pulse selectivity of $1 - e^{\frac{-\chi^2}{2\sigma^2}} > 0.99 $. This is essentially a CNOT~\cite{Johnson:2010gu} operation, with the qubit as the target, conditioned on the photon number in the cavity. This process can be described by
\begin{equation}\label{cnot}
	\ket{\Psi_{cavity}} \otimes \ket{\Psi_{qubit}}  = \sum_{n=0}^\infty p_n \ket{n} \otimes \ket{g} \xrightarrow{X_\pi^m} \sum_{\substack{n=0 \\ n \neq m}}^\infty p_n \ket{n} \otimes \ket{g} + p_m \ket{m} \otimes \ket{e}.
\end{equation}
Where the state of the cavity is described as a sum over the Fock states $\ket{n}$ with $p_n$ the complex amplitude.
This procedure entangles the qubit and cavity thus a measurement of the qubit state will project the cavity onto the $m^{th}$ photon fock state. The probability of finding the qubit in the excited state is then given by $p_m$ which is exactly the probability of finding the cavity in the state $\ket{m}$.

\section{Determining the cavity anharmonicity}

\begin{figure*}[h] 
   \includegraphics[width=183mm]{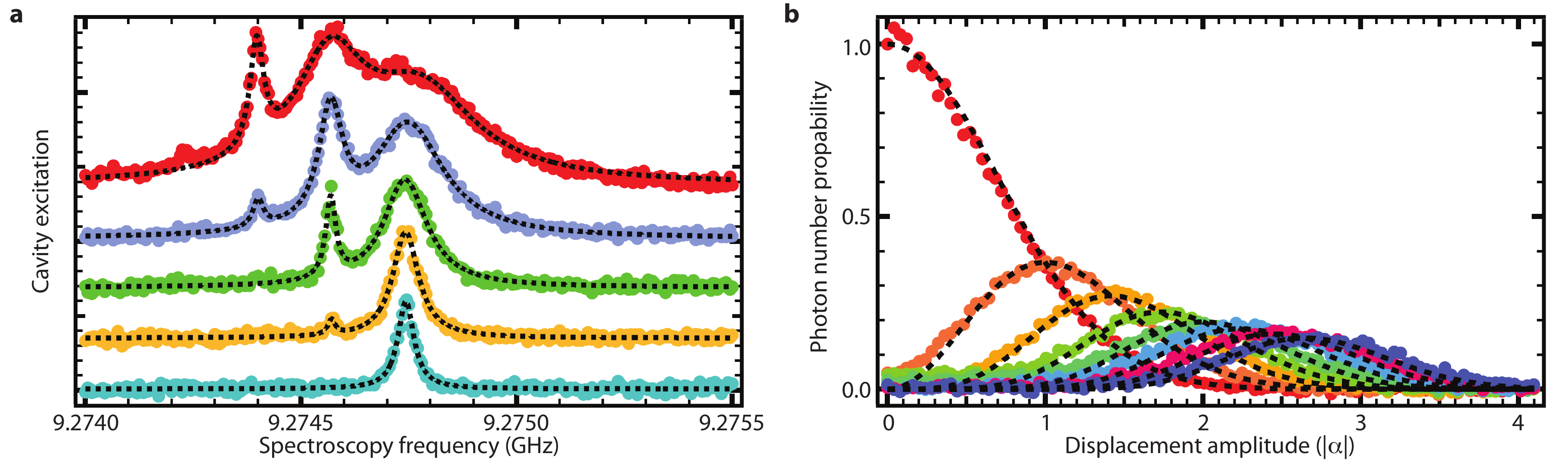} 
   \caption{\textbf{Cavity spectroscopy and displacement calibration.} \textbf{a}, Cavity spectroscopy for varying spectroscopy power. The cavity excitation is measured by mapping the population left in the ground state to the qubit using a photon number state selective $\pi$ pulse. For the lowest power (\textcolor{BlueGreen}{$\bullet$}) the cavity is only excited on the $\ket{0} \rightarrow \ket{1}$ transition. As the power is increased ($\textcolor{BlueGreen}{\bullet}\rightarrow\textcolor{Dandelion}{\bullet} \rightarrow \textcolor{ForestGreen} {\bullet} \rightarrow \textcolor{CadetBlue}{\bullet}\rightarrow \textcolor{Red}{\bullet}$) different transitions appear in the spectroscopy. These transitions are n-photon transition from $\ket{0}$ to $\ket{n}$ with $n=1, 2, 3$ labeled from the right peak to left. The separation between the peaks is given by $0.5\, K/2\pi= 163$~kHz. The dashed lines are fit to the data using a multipeak Lorentzian function.  \textbf{b}, Photon number probability of the Fock states $n=0\dots7$ ($\textcolor{Red}{\bullet}, \textcolor{RedOrange} {\bullet}, \textcolor{BurntOrange}{\bullet}, \textcolor{LimeGreen}{\bullet},\textcolor{JungleGreen}{\bullet},\textcolor{RoyalBlue}{\bullet},\textcolor{RubineRed}{\bullet},\textcolor{BlueViolet}{\bullet}$) for a 10~ns displacement with amplitude $|\alpha|$ of the cavity ground state. The axes were scaled by fitting a Poisson distribution with two free fit parameters to all seven Fock state populations simultaneously. The dashed lines are given by a Poisson distribution $P(|\alpha|) = |\alpha|^{2n} e^{-|\alpha|^2}/n!$ for the Fock state n.}
   \label{fig:cavityspec}
\end{figure*}

The cavity anharmonicity is measured by spectroscopy using a 50 $\mu$s pulse to excite the cavity. The cavity excitation is measured by mapping the population left in the ground state to the qubit using a photon number state selective $\pi$ pulse. The $\pi$ pulse on the qubit will be successfull only when the cavity is in the ground state after the spectroscopy pulse. Spectroscopy on the cavity for a varying spectroscopy power is shown in Fig.\ref{fig:cavityspec}a. For the lowest power the cavity is only excited on the $\ket{0} \rightarrow \ket{1}$ transition at a frequency of 9.2747~GHz. As the power is increased different transitions appear in the spectroscopy. These transitions are multi-photon transitions from $\ket{0}$ to $\ket{n}$ with $n=2, 3$ which means the separation between the peaks is given by $0.5\,K/2\pi=$ 163 kHz, the Kerr nonlinearity of the cavity. The data were fit using a sum of Lorentzian peaks.

\section{Cavity displacement \& displacement calibration}

The cavity is displaced using a 10~ns pulse resonant with the $\ket{0} \rightarrow \ket{1}$ transition of the cavity creating a photon number probability distribution given by a coherent state. We can measure the probability of finding the Fock state $\ket{n}$ by applying a photon number state selective $\pi$ pulse to the qubit and measure the qubit excitation. The displacement voltage $\epsilon$ applied to the cavity is calibrated in units of the average photon number in the cavity $\bar{n}$, or rather the corresponding displacement amplitude $|\alpha| = \sqrt{\bar{n}}$, by fitting a Poisson distribution $P_n(\epsilon) = (\frac{\epsilon}{\Delta\epsilon})^{2n} e^{(\epsilon/\Delta\epsilon)^2}/n!$, with the normalization constant $\Delta\epsilon$, to all measured photon number probabilities simultaneously. The probability of finding the Fock states $n=0\dots7$  for a displacement with amplitude $|\alpha|$ of the ground state can be seen in Fig.\ref{fig:cavityspec}b. The probabilities closely follow the expected Poisson distribution. This demonstrates that it is possible to create a coherent state despite the cavity non-linearity as long as the displacement pulse has a large enough spectral width.

\section{Excited state qubit population and negative values in the Hussimi Q-function.}

We found, that the qubit has a 10\% excited state population which has been observed in other 3D superconducting qubits~\cite{Corcoles:2011is,Sears:2012uca} but is not yet completely understood. In our case this excited state qubit population leads to negative values in the Hussimi Q-function, as can be seen in Fig.~3a of the main text. These negative values are a result of the measurement procedure as $Q_0(\alpha)=0$ is given by the readout voltage corresponding to the steady state population of the qubit. When the qubit excited state population is rotated to the ground state the readout voltage will be lower than for $Q_0(\alpha)=0$. This rotation becomes possible as the resonator state for the qubit in the excited state evolves at a rate different by $\chi$ compared to the ground state qubit resonator. The two coherent states will separate in phase space as can be seen in  Fig.~3a of the main text. The evolution of this coherent state correlated with an excited state qubit will closely mirror the ground state population up to an additional rotation with $\chi$. This means that at $T_{\mathrm{rev}}/q$ we will also generated q-component cat states which have negative readout voltages. This will reduce the amplitudes and coherent fringes of the observed cat states as the amplitudes of our $Q_n$ measurements are scaled to the amplitude of the coherent state of Fig. 3a in the main text. Effectively, the excited state population will  lead to a reduction of the fidelity for the measured cat states.

\section{Cavity state tomography}
\begin{figure*}[h] 
   \includegraphics[width=183mm]{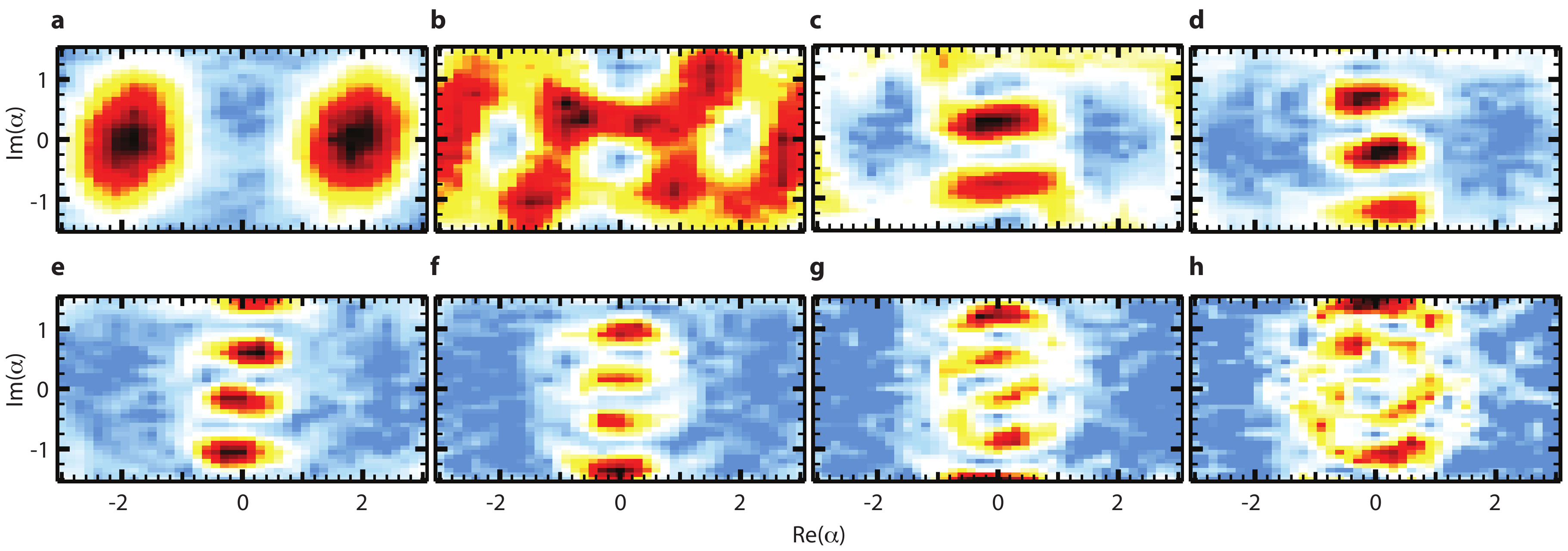} 
   \caption{\textbf{\boldmath{$Q_n(\alpha)$} probability distributions.} \textbf{a}-\textbf{h} show the measured projections on the Fock states with $n=0-7$  for the two-component Schr\"odinger cat state. Each $Q_n(\alpha)$ was measured at 441 different displacements $\alpha$. The different $Q_n(\alpha)$  functions were measured by conditioning the qubit $\pi$ pulse on the photon number in the cavity.}
         \label{fig:Qn}
\end{figure*}

Using the photon number state selective pulse to project onto the $\mathrm{n^{th}}$ photon Fock state, along with a preceding cavity displacement, we can measure (see Fig.~\ref{fig:Qn}):

\begin{equation}\label{husimiQ}
 Q_n(\alpha_m) = \frac{1}{\pi} \langle n | D(-\alpha_m) \rho D(\alpha_m) | n \rangle\\
\end{equation}
where $n$ is the fock state projection; $\alpha_m$,the tomography displacement; and $\rho$, the cavity density matrix. The $Q_n$ are normalized probability distributions. Assuming a truncated photon basis, we are also able to directly reconstruct the cavity Wigner function~\cite{Hofheinz:2009tt,Leibfried:1996}:
\begin{equation}\label{wigner}
	\begin{split}
		W(\alpha_m)& = \frac{2}{\pi} \; \mathrm{Tr}( D(-\alpha_m) \rho D(\alpha_m) P)\\
			     & =  \frac{2}{\pi} \sum_n (-1)^n \langle n | D(-\alpha_m) \rho D(\alpha_m) | n \rangle\\
			     & =  2 \sum_n (-1)^n Q_n(\alpha_m)
	\end{split}
\end{equation}
where $P$ is the photon number parity operator. A more effective method for determining the Wigner function of the resonator is to reconstruct its density matrix. Realizing that the $Q_n$ are a set of linear equations, we can rewrite Eqn.~\ref{husimiQ} as:
\begin{equation}\label{linearEq}
	\begin{split}
		Q_n(\alpha_m) = \frac{1}{\pi}  \sum_{ij} M_{nmij} \rho_{ij}\\
	\end{split}
\end{equation}
where $M_{nmij} = \langle n| D(-\alpha_m) | i \rangle \langle j | D(\alpha_m) | n \rangle$  and $\rho_{ij}=\bra{i}\rho\ket{j}$. Measuring the distributions, $Q_n(\alpha_m)$, with known photon projections n and displacements $\alpha_m$, we can perform a least squares regression to determine the cavity state, $\rho$. For this experiment, we measured $441$ displacements each for projections onto the Fock states with $0$ to $7$ photons, 3528 measurements in total with 1000 averages per measurement point for each regression. We perform the linear regression with a priori assumptions that the cavity density matrix is Hermitian, of trace one, positive semi-definite and having a trunctated basis of ten photons. This reconstructed density matrix is then used to calculate state fidelities and plot Wigner functions.

\bibliographystyle{naturemag}
\bibliography{biblio}